\documentclass{aa}
\usepackage{txfonts}
\usepackage{graphicx}
\usepackage{natbib}
\bibpunct{(}{)}{;}{a}{}{,}

\begin{document}

\title{First spectra of the W UMa system V524 Monocerotis\thanks{Based 
on observations made with FEROS at the
ESO/MPI 2.2m telescope at ESO, La Silla, Chile}}
\subtitle{}

\author{T. H. Dall
	\and
	L. Schmidtobreick
}

\institute{
	European Southern Observatory, Casilla 19001, Santiago 19, Chile
}

\date{Received / Accepted }

\abstract{
We present the first high-resolution spectra of the W\,UMa contact binary
\object{V524 Mon}. The spectra of the two components are very similar, resembling
a G5 and a K0. We find the radial velocities and rotational velocities consistent with corotation. We estimate the radii and the masses
and derive a mass ratio $M_2/M_1 = 2.1$. We confirm that V524\,Mon is a W--type contact system,
likely enclosed in a common convective envelope, as found
by \citet{samec+loflin2003}.  We do not find evidence for the expected level of emission in
the chromospheric \ion{Ca}{ii} H and K lines, neither H$\alpha$, indicating 
that the magnetic activity is much weaker than expected or that other processes are hampering chromospheric emission.
}

\titlerunning{The W\,UMa system V524 Monocerotis}
\maketitle

\section{Introduction}
The W UMa variables are eclipsing contact biaries with components of spectral types F -- K, with periods in the range 0.2 to 1.5 days,
although the distribution is strongly peaked between 0.25--0.6 days \citep{rucinski1994,rucinski1998}.
As the components of these systems are rapid rotators, they normally show strong chromospheric and coronal emission \citep{rucinski+1985} obeying
a period-activity relation, where the activity increases towards shorter periods.
However, for the very fast rotators a saturation of the magnetic activity is reached. In the X-ray emitting corona, this is
the result of a decreased coronal filling factor, caused by the partial covering of the two components by an envelope
of matter flowing from the seconday to the primary and back, partly covering the surfaces and suppressing long-lived magnetic
structures \citep{stepien+2001}.  In the chromosphere, the flow of matter should not seriously dampen the magnetic activity. 
Here the main problem is measuring the emission features, because of the extreme rotational broadening and the underlying continuum. 
For this reason the UV range where there is little continuum is best suited for chromospheric studies, although for the late spectral types
the \ion{Ca}{ii} H and K lines should be accessible as well.

Recently, \citet{samec+loflin2003} presented precise UBVRI light curves of the overlooked
W\,UMa system \object{V524\,Monocerotis} (\object{GSC 00153-01410}, B\,=\,14.40). From their own data and the literature, they found
an orbital period P~=~0.283616\,d, hence a short period system which is expected to be very active. They presented a quadratic ephemeris,
with a decreasing period, which, according to \citeauthor{samec+loflin2003}, could indicate that V524 Mon
is losing angular momentum, presumably due to magnetic braking. 
While in a previous analysis, \citet{hoffmann1981} suggested the lightcurve to
be of A--type, \citeauthor{samec+loflin2003} find W--type lightcurves which
indicate ``the presence of heavy, saturated magnetic activity''. Their model had a mass ratio $M_2/M_1 = 1.84$, fill-out 4.5\%
and a hot spot on the primary.

In this paper we present the first high-resolution spectra of \object{V524\,Mon}. The spectra confirm the model of
\citeauthor{samec+loflin2003} based on derived rotational and radial velocities, and the inferred spectral types. However, 
the spectra also show less than expected chromospheric emission.

\section{Observations}
We have obtained three high-resolution (R$\sim$48000)
spectra of V524 Mon, using FEROS at the ESO/MPI-2.20m telescope at La Silla, Chile.
The spectra all have  exposure times of 1800\,s and were acquired during
on-the-fiber guiding tests on faint targets. The mid-exposure times are
HJD\,2453018.69848,
2453018.72169 and 2453018.74310, corresponding to phases 0.12, 0.20 and 0.28 respectively, using the
quadratic ephemeris of \citeauthor{samec+loflin2003}. The linear ephemeris give indistinguishable phases.

Standard data reduction was performed with MIDAS
including bias and flatfield correction, order extraction and wavelength calibration.
The spectra have a FWHM resolution of 0.15\AA\ and cover the 
range 3800--9000\AA . FEROS is not the optimum choice for such faint targets, 
so the S/N is $\sim 40$ around H$\alpha$, declining to 10 -- 15 near the \ion{Ca}{ii} resonance lines.
  No velocity or spectrophotometric standards were observed during the night.

\section{Analysis and discussion}
Evident in all three spectra are absorption features of the two stellar components from 
the Balmer lines, \ion{Ca}{ii}~H+K, \ion{Ca}{i}\,($\lambda$\,4227\,\AA), \ion{Fe}{i}\,($\lambda$\,4383\,\AA) 
\ion{Mg}{i}\,Eb and \ion{Na}{i}\,D, 
all rotationally broadened. Also present are narrow
Na\,D interstellar absorption lines.

\subsection{Binary parameters}
As expected for a W\,UMa--system, there is a very high degree of similarity 
between the spectra of the two components, which indicates that the stars are 
indeed enclosed in a common convective envelope.
  The strong rotational broadening of the
stellar lines makes a  spectral classification of the two components quite difficult.
Nevertheless, all recognisable lines which are typical of late-G to early-K are present 
in both stars. They also have
approximately the same relative strengths, except for the Balmer lines 
which are clearly stronger for
component~1 than for component~2 \citep[labelling follows][]{samec+loflin2003}.
On the basis of our spectra we assign a formal spectral type of G5 for the hotter component (1) and K0 
for the companion (2).
Because of the broadening and the low S/N we estimate a few subclasses of uncertainty in the spectral classification.
The interpretation would be as follows: the hotter component 
(1) is brighter and shows stronger Balmer lines. 
The cooler component (2) shows intrinsically stronger \ion{Ca}{i} and \ion{Fe}{i} but weaker Balmer 
lines. Since it is fainter, the Fe and Ca lines appear to be of equal strength, 
while the Balmer lines appear very weak. 
This is in full agreement with the model of Samec \& Loflin, where the 
larger but fainter star (2) has the cooler surface.

\begin{figure*}
\includegraphics[width=17cm]{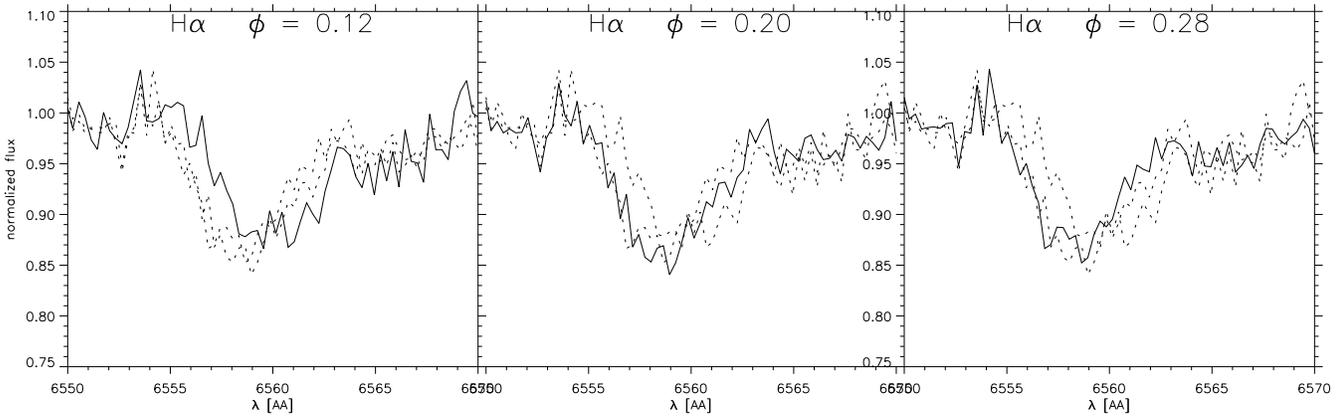} 
	\caption{The H$\alpha$ line region in each spectrum, rebinned to suppress the noise. At each 
phase, the two other spectra are over-plotted for reference.}
	\label{fig:halpha}
\end{figure*}

\begin{figure}
        \resizebox{\hsize}{!}{\includegraphics{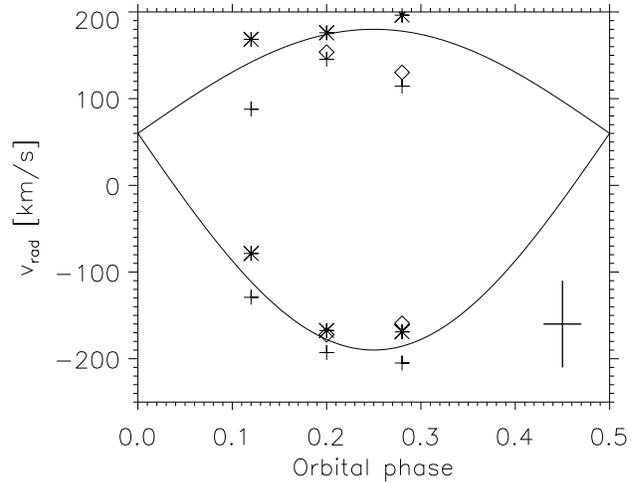}}
\caption{Measured radial velocities and simple sinusoidal 
fits to the data. The cross in the lower right
corner indicate typical error bars. *: Fe; +: H$\alpha$; $\diamond$: Ca.}
	\label{fig:radvel}
\end{figure}

In Fig.~\ref{fig:halpha}, for all three spectra a region centred on H$\alpha$ is plotted, 
showing clearly the change of line position and shape with orbital phase.
From the lines of \ion{Fe}{i} ($\lambda$\,4383.544\,\AA) and \ion{Ca}{i} ($\lambda$\,4226.728\,\AA), as well as
from H$\alpha$, we estimate the radial velocities as a function of phase. The 
values are given in Table~\ref{tab:radvel} and plotted
in Fig.~\ref{fig:radvel}, together with a simple sinusoidal fit, i.e. assuming a circular orbit. 

\begin{table}
\centering
\caption{Radial velocity as function of phase measured in different absorption lines}\label{tab:radvel}
\begin{tabular}{lrrr} \hline
Component  & \multicolumn{3}{c}{$v_{\mathrm{rad}}$ [km\,s$^{-1}$]}  \\
... @ phase & \ion{Ca}{i}\,(4227)  &  H$\alpha$ & \ion{Fe}{i}\,(4383) \\ \hline
{\bf 1} @ 0.12 &        & -129.2 &  -78.6 \\
{\bf 1} @ 0.20 & -172.2 & -192.9 & -167.5 \\
{\bf 1} @ 0.28 & -159.1 & -205.0 & -169.0 \\ \hline
{\bf 2} @ 0.12 &        &  87.8  & 168.4 \\
{\bf 2} @ 0.20 & 153.7  & 145.4  & 176.0 \\
{\bf 2} @ 0.28 & 130.2  & 114.4  & 196.3 \\ \hline 
\end{tabular}
\end{table}

We find a best fit for a system radial velocity of 60\,km\,s$^{-1}$ and
amplitudes of 250\,km\,s$^{-1}$ and 120\,km\,s$^{-1}$ for the two components. Note that the component 
with the strong Balmer lines has the higher amplitude and is hence the less massive star. From this we derive a
mass ratio $M_2/M_1 = 2.1\pm0.4$, consistent with the findings of \citeauthor{samec+loflin2003}, who from their model fit find
a mass ratio of 1.84.
Note that our velocity errors 
are rather high due to the extreme broadening of the lines, and that 
the fit is done over only three points along the orbit, although covering the times of maximum radial velocity separation. 
Hence, we expect the listed radial velocities
to be accurate only to within $\pm 40$\,km\,s$^{-1}$. 

Using our derived radial velocities together with the period 
$P= 0.2836$\,d and the inclination $i = 79\fdg1$
derived by \citeauthor{samec+loflin2003}, we compute the masses of the two 
components as $M_1 = 0.5(3) \rm M_\odot$ and $M_2 = 1.1(6)\rm M_\odot$.
Both masses are inconsistent with the corresponding stellar type as can be
expected for components of a system with a common convective envelope.
However, the mass ratio together with the spectral types of the components,
i.e. the fact that the star of later type has the higher mass, does indicate 
that component 2 is an already
evolved star which reached contact during its evolution off the main sequence,
and has lost some part of its mass to the common envelope
\citep[see e.g.][]{iben+livio1993}.

From the line-broadening, we estimate 
the rotational velocities to be 220\,km\,s$^{-1}$ and 240\,km\,s$^{-1}$ 
for components 1 and 2 respectively. 
This gives a radius difference of only $\sim 10$\% assuming the stars 
are co-rotating. For the individual radii we then get
$R_1=8.7(4)\times 10^5\rm km = 1.25 R_\odot$ and 
$R_2=9.4(4)\times 10^5\rm km = 1.35 R_\odot$,
using an inclination of $79\fdg1$.

In Table~\ref{tab:summary} we summarise the derived parameters for V524\,Mon.
\begin{table}
\label{tab:summary}
\centering
\caption{The derived parameters for V524 Mon. See text for error estimates.}
\begin{tabular}{l|rr} \hline
   & Component 1 & Component 2 \\ \hline
Sp.Type & G5 & K0 \\
Mass (M$_\odot$) & 0.5 & 1.1 \\
Radius (R$_\odot$) & 1.25 & 1.35 \\
RV ampl. [km\,s$^{-1}$] & 250 & 120 \\
$v_\mathrm{rot}$ [km\,s$^{-1}$] & 220 & 240 \\ \hline
\end{tabular}
\end{table}

\subsection{Magnetic activity}
The efficiency of the dynamo and hence the activity level is usually described in terms of the 
Rossby number, $R_0 = P \tau_c^{-1}$, where $P$ is the rotation period and $\tau_c$ is
the convective turnover time. It has been established empirically \citep[][and subsequent studies]{noyes+1984} 
that the activity of single stars scales
with $R_0^{-1}$. 
At very low Rossby number (very fast rotation) a saturation sets in: the activity stays roughly constant as the 
rotation period decreases. This is 
thought to be linked to the complete filling of the stellar surface by active regions. 

\begin{figure*}
\includegraphics[width=17cm]{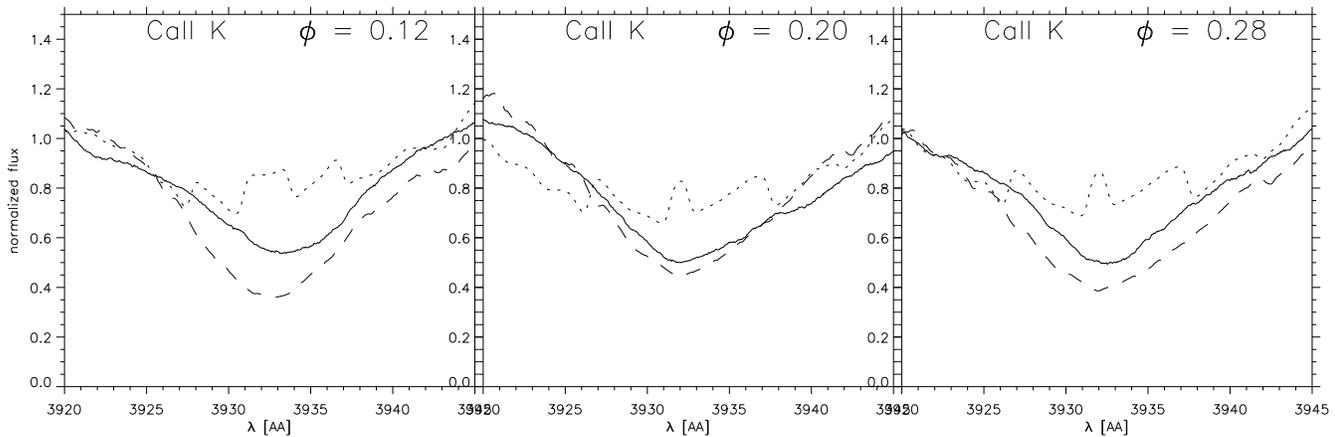}
	\caption{The \ion{Ca}{ii} K line at the three phases (solid line) smoothed to suppress the noise. 
The dashed line is constructed from a solar
spectrum and the dotted line from a spectrum of a well-known active star --- see text for discussion.}
	\label{fig:caII}
\end{figure*}
The late-type (W--type) W\,UMa systems all have very high activity levels, as has been 
confirmed by several studies \citep[e.g.,][]{cruddace+dupree1984,rucinski+1985,stepien+2001}. However, because of the extreme rotational 
broadening, the usual optical
chromospheric activity indicators, the \ion{Ca}{ii} emission cores, are very difficult to observe. 

In Fig.~\ref{fig:caII} we show our three
spectra of V524\,Mon centered on the \ion{Ca}{ii}\,K line.
 The dashed line
shown on the figure was constructed from a spectrum of the \object{Sun} in the following way: 
Two  versions of the same solar spectrum were added, one 
of them shifted in wavelength according to the orbital phase. 
The result was smoothed to wash out the fainter lines, mimicking the effects of rotation, 
and then plotted on top of the similarly smoothed V524 Mon spectrum. The upper dotted curve was constructed in the same way 
but with a spectrum of  \object{HD\,82558} (LQ\,Hya, K0Ve), an active star of the BY\,Dra type
which shows strong \ion{Ca}{ii} emission lines. Both spectra were obtained by Dall (2004; unpublished) 
using the same setup as for V524\,Mon.
Since the spectral types are similar, 
and since the S/N is too low to allow a reliable fit, this approach is
a valid way to look for excess emission in the rotationally broadened line profiles. 
Neither the Sun nor HD\,82558 provides a satisfactory fit to the spectra of V524\,Mon. We have chosen to match as
closely as possible the far wings of the lines in a conservative way, i.e. chosen the lowest possible ``safe'' positioning of the 
HD\,82558 template and the highest possible one for the solar  template in order
to minimize the fitting bias.
Even so, the spectrum of V524\,Mon does not show the level of emission fill-in  seen in the HD\,82558 template, but rather resembles
the line shape of the solar case.

We will now justify the choice of HD\,82558 as a reference for the expected \ion{Ca}{ii} emission. It is generally accepted 
that all active binaries obey the flux-flux scaling relations established 
from the study of single stars \citep[see e.g.][]{schrijver+zwaan2000,messina+2001}.
Moreover, \citet{rucinski1985} notes that the late-type W\,UMa systems have activity levels
expected by  extrapolating from moderately rotating G-K  stars to large values of $R_0^{-1}$ (i.e., fast rotation).
Hence, we can estimate the expected emission from the scaling relation between the transition line 
and chromospheric line fluxes presented by \citet{montes+1996} 
for active binaries. Using a typical \ion{C}{iv} line flux listed by \citet{vilhu+walter1987} for W\,UMa systems, we find
for the \ion{Ca}{ii}\,K line F$_\mathrm{S}$(\ion{Ca}{ii}\,K)$\,\sim\,10^7$\,erg\,cm$^{-2}$\,s$^{-1}$\,\AA$^{-1}$, 
i.e. roughly equal to the flux from
the most active systems of the BY\,Dra type, hence the comparison with HD\,82558 is justified.   

It is clear from Fig.~\ref{fig:halpha} that the H$\alpha$ line appears in absorption, which is another indication of the
lack of strong activity.  We have not attempted to estimate whether some fill-in is present in H$\alpha$. Because of the
strong temperature dependence of the Balmer lines, any such estimates would be higly uncertain given the uncertainties
of spectral type and temperature.

It is clear from Fig.~\ref{fig:caII}
that any emission present in the core of the \ion{Ca}{ii}\,K line is much weaker than expected.
In the following we will discuss the possible causes for the lack of strong
\ion{Ca}{ii} emission in the spectra of V524\,Mon.

All studies of single stars indicate that the activity is well correlated spatially, from the corona to
the photosphere \citep{messina+2001}, meaning that the flux-flux relationships are valid. Up to the present, there have been no indications
that this was not the case for binaries as well. 
Furthermore, according to the study by \citet{montes+1996}, binaries tend to be overactive relative to single stars, although still following a $R_0^{-1}$ relation.
Saturation is thought to have occured
for all W UMa systems due to the extreme rotation.

Recently \citet{stepien+2001} argued that the asymmetric distribution of coronal X-ray 
activity on W UMa stars \citep{brickhouse+dupree1998,gondoin2004} is due to the attenuation of long-lived 
magnetic regions by mass flows. In this picture, the mass flow covers the whole surface of the secondary and a broad equatorial
band on the primary, drawing  magnetic loops back into the photosphere before they become large enough to reach the corona, effectively reducing the 
coronal filling factor. 
Since we have only sampled a fraction of the orbit of V524\,Mon, one explanation of the missing emission might be that the
chromospheric activity has a highly uneven spatial distribution, resembling the situation for the corona, which would 
mean that the flux-flux relationships are not valid.
However, several studies seem to  indicate that the chromospheric activity is rather uniform over the common
convective envelope and that the flux in a given emission line is constant over the orbit \citep[e.g.,][]{rucinski+1985,vilhu+walter1987}.
Moreover, the inferred asymmetry of the corona may be biased by heavy X-ray flaring activity, hence the 
amount of asymmetry in the corona is not a settled issue.

If the two stars are not yet in contact, then the chromospheric emission pattern may very well be asymmetric as observed by
\citet{duemmler+2003}. However, the excellent fit to the W--type lightcurves of V524\,Mon, and the indication of corotation argues
strongly against this possibility.

Another possibility is  that the definition of the Rossby number, and indeed the concept of the usual differential-rotation fed dynamo
may be quite meaningless in the close, rapidly rotating environment of a  contact system, calling for a revision of the current picture of
dynamo action in very rapidly rotating stars, and possibly also incorporating the unique effects of a close binary.  
It seems, though, that there exist activity cycles in W\,UMa systems \citep[see][for a recent discussion]{yang+liu2003}. These have
periods comparable to the solar case, hence should we by chance have observed V524\,Mon 
during a ``quiet'' state, then there is a good chance 
that quick follow-up observations will find it still in this state.

Finally, the problem may lie with the \ion{Ca}{ii} lines themselves. These have a fairly low formation temperature compared
to most other (UV) chromospheric emission lines, and hence if the interface between the photosphere and the chromosphere 
is quite different 
to the solar case, there may exist physical conditions that inhibit the formation of these lines. 
However, the emission cores of the \ion{Mg}{ii} h and k lines,
which are formed at approximately the same temperatures, are generally observed in W UMa stars.

\section{Conclusion}
Comparing our findings with the model presented by \citet{samec+loflin2003}, we 
find good overall agreement.  We determine formal spectral types of G5 for the hotter, less massive
component and K0 for the larger component. Thus we confirm their interpretation
of V524\,Mon being a W\,UMa contact binary of W--type and refute Hoffmann's 
classification, also confirming that photometric light-curve fitting and modelling \citep{bradstreet1992,wilson1994} does 
produce excellent and reliable results.

However,  all such short-period late-type systems have very high activity levels, as has been 
confirmed by several studies, but our study
does not bring forth clear evidence of  strong chromospheric activity. On the contrary, there seems to
be an emission-deficiency in the cores of \ion{Ca}{ii} H and K with respect to the expected activity level.

We suggest that the discrepancy could be due to an incomplete understanding of the influence of
very high rotation rates on the structure of the convection zone and the differential rotation, and hence on the dynamo mechanism.
We also believe that a better understanding of the influence of close binarity on the 
generation and morphology of magnetic fields is necessary.  Alternatively, 
we suggest that the conditions in the chromosphere of the star may suppress the \ion{Ca}{ii} H+K emission, although we note
that the similar \ion{Mg}{ii} h+k lines follows the established flux-flux relations in other W UMa stars.

It seems that it is at present not possible, or at least very difficult, to use the \ion{Ca}{ii} emission as an indicator of 
chromospheric activity in  contact systems, whether or not this is due to intrinsic effects.

More observations are needed to clarify the activity status of this supposedly very active system.

\begin{acknowledgements}
We are very grateful to V.\,Andretta and S.\,Bagnulo for helpful discussions and suggestions to improve the manuscript.
This research has made use of the SIMBAD database, operated at CDS, Strasbourg, France.
\end{acknowledgements}

\bibliographystyle{bibtex/aa}
\bibliography{1300refs.bib}

\end{document}